\documentclass[us letters]{article}
\usepackage{amsmath,graphicx,amssymb,delarray,subfigure,verbatim, bbding}
\usepackage{multirow}
\usepackage{interspeech2021}

\title{A Simultaneous Denoising and Dereverberation Framework with Target Decoupling}
\name{Andong Li$^{1, 2}$, Wenzhe Liu$^{1,2}$, Xiaoxue Luo$^{1,2}$, Guochen Yu$^{1, 3}$, Chengshi Zheng$^{1,2}$\thanks{Chengshi Zheng is the corresponding author.}, Xiaodong Li$^{1,2}$}
\address{
	$^1$Key Laboratory of Noise and Vibration Research, Institute of Acoustics, Chinese Academy of Sciences, Beijing, China\\
	$^2$University of Chinese Academy of Sciences, Beijing, China\\
	$^3$Communication
	University of China, Beijing, China}
\email{ \{liandong, liuwenzhe, luoxiaoxue, cszheng, lxd\}@mail.ioa.ac.cn, yuguochen@cuc.edu.cn}

\begin{document}

\maketitle
\begin{abstract}
Background noise and room reverberation are regarded as two major factors to degrade the subjective speech quality. In this paper, we propose an integrated framework to address simultaneous denoising and dereverberation under complicated scenario environments. It adopts a chain optimization strategy and designs four sub-stages accordingly. In the first two stages, we decouple the multi-task learning w.r.t. complex spectrum into magnitude and phase, and only implement noise and reverberation removal in the magnitude domain. Based on the estimated priors above, we further polish the spectrum in the third stage, where both magnitude and phase information are explicitly repaired with the residual learning. Due to the data mismatch and nonlinear effect of DNNs, the residual noise often exists in the DNN-processed spectrum. To resolve the problem, we adopt a light-weight algorithm as the post-processing module to capture and suppress the residual noise in the non-active regions. In the Interspeech 2021 Deep Noise Suppression (DNS) Challenge, our submitted system ranked top-1 for the real-time track in terms of Mean Opinion Score (MOS) with ITU-T P.835 framework.
\end{abstract}
\vspace{-0.2cm}
\section{Introduction}
\vspace{-0.2cm}
As an indispensable front-end technique in many speech-related tasks like telecommunication and automatic speech recognition (ASR) systems, speech enhancement (SE) has been widely investigated over the past several decades~{\cite{loizou2013speech}}. It aims to extract the speech target from the degraded speech and improve the speech quality when background noise and room reverberation are available. Recent years have witnessed the unprecedented development of deep neural networks (DNNs) and a multitude of DNN-based SE algorithms have been proposed, which are reported to notably surpass traditional signal-processing-based methods~{\cite{wang2014training, pascual2017segan}}.

For a long time, DNN-based SE algorithms attempt to recover the speech in the time-frequency (T-F) domain. Due to the unstructured characteristic of the phase distribution, they only focus on magnitude estimation while leaving the phase unaltered~{\cite{wang2014training, xu2014regression,strake2020fully}}. Recently, the importance of phase begin to be emphasized from the perception aspect~{\cite{paliwal2011importance}}, and a multitude of phase-aware algorithms are proposed since then~{\cite{williamson2015complex, tan2019learning}}. For these approaches, the multi-target learning is usually considered, which optimizes magnitude and phase information simultaneously, \emph{e.g.}, the complex ratio masks (CRMs)~{\cite{williamson2015complex}} and direct complex spectral coefficients~{\cite{tan2019learning}}. 

More recently, time-domain-based approaches begin to gain renaissance, where both feature domain transform and denoising process are implicitly integrated within one network~{\cite{pascual2017segan, fu2018end, luo2019conv, strake2020speech}}. Despite the impressive performance of the time-domain-based algorithms, they usually suffer from several drawbacks. For one thing, different from ideal (i)STFT, as the domain transform coefficients are learned in a data-driven manner, they usually exhibit relatively poor generalization capability toward the real acoustic environments. For another, it has been illustrated that the noise and speech components tend to be more separately distinguishable in the T-F domain~{\cite{yin2020phasen}}. Moreover, we notice that T-F domain-based approaches usually exhibit better Mean Opinion Score (MOS) in the previous Deep Noise Suppression (DNS) Challenge{\footnote{https://www.microsoft.com/en-us/research/academic-program/ deep-noise-suppression-challenge-icassp-2021/}}. As a result, we still focus on interference suppression in the T-F domain in this study.

Inspired by the curriculum learning concept~{\cite{bengio2009curriculum}}, multi-stage SE begin to thrive in the recent several years~{\cite{strake2020speech, li2020speech, li2020two, li2021icassp, gao2018densely}}. Different from the previous one-step paradigm, the whole mapping process is decomposed into several sub-stages and the network is coerced to learn following either ``from easy to hard'' or ``from hard to easy'' training paradigm~{\cite{fayek2020progressive}}. As a consequence, the abundant prior information can be utilized to facilitate the subsequent optimization. In our preliminary study, we propose a two-stage paradigm for noise suppression~{\cite{li2020two}}. In this paper, we further extend the framework for simultaneous \textbf{S}peech \textbf{D}enoising and \textbf{D}ereverberation, which is called \textbf{SDD-Net}. The whole framework takes a chain-optimization strategy and can be divided into four parts. In the first two parts, the optimization \emph{w.r.t.} denoising and dereverberation are serially implemented in the magnitude domain, \emph{i.e.}, we decouple the original complex multi-target optimization into magnitude and phase, and only focus on magnitude estimation. Afterward, the phase information can be effectively modified based on previous results by introducing a global residual connection. The rationale is that there exists implicit relation between magnitude and phase and superb magnitude estimation can profit better recovery for phase~{\cite{yin2020phasen, 9242279}}.

Due to the highly nonlinear characteristic of DNNs, some residual artifacts inevitably exist after the process. Generally, these artifacts are audible and noncontinuous, which heavily hinder the subjective quality. To ameliorate the issue, we take a post-processing module to remove the remaining distortion~{\cite{li2021icassp}}, which proves significant to improve the MOS.

The rest of the paper is organized as follows. In Section~{\ref{problem-formulation}}, we formulate the problem. The overall diagram is illustrated in Section~{\ref{system-overview}}. We present the experimental setup in Section~{\ref{experiments}}. The results and analysis are given in Section~{\ref{results-and-analysis}}, and some conclusions are drawn in Section~{\ref{conclusion}}.
\vspace{-0.2cm}
\section{Problem formulation}
\label{problem-formulation}
\vspace{-0.2cm}
In the time domain, let $s\left(t\right)$, $h\left(t\right)$, and $n\left(t\right)$ denote the anechoic, room impulse response (RIR) and background noise, respectively, then the received speech $y\left(t\right)$ can be written as:
\begin{equation}
\label{eqn1}
y\left(t\right) = s\left(t\right) * h\left(t\right) + n\left(t\right) = x\left(t\right) + r\left(t\right) + n\left(t\right),
\end{equation}
where $*$ denotes the convolution operator. $x\left(t\right)$ and $r\left(t\right)$ refer to the direct sound plus early reflections, and late reverberation. With STFT, Eq.~{(\ref{eqn1})} can be converted into the T-F domain:
\begin{equation}
\label{eqn2}
Y_{l, m} = X_{l, m} + R_{l, m} + N_{l, m},
\end{equation}
where $Y_{l, m}$, $X_{l, m}$, $R_{l, m}$, and $N_{l, m}$ are the corresponding variables with frame index of $l$ and frequency index of $m$. For legibility, we will omit the subscript if no conflict arises. In this paper, we aim to remove the noise and late reverberation, and the target estimation probability $p\left(X|Y\right)$ can be factorized by using the probabilistic chain rule, given as follows:
\begin{equation}
\label{eqn3}
p\left(X|Y\right) = p\left(N|Y\right)p\left(R|Y, N\right)p\left(X|Y, R, N\right),
\end{equation}

\begin{figure*}[t]
	\centering
	\centerline{\includegraphics[width=1.50\columnwidth]{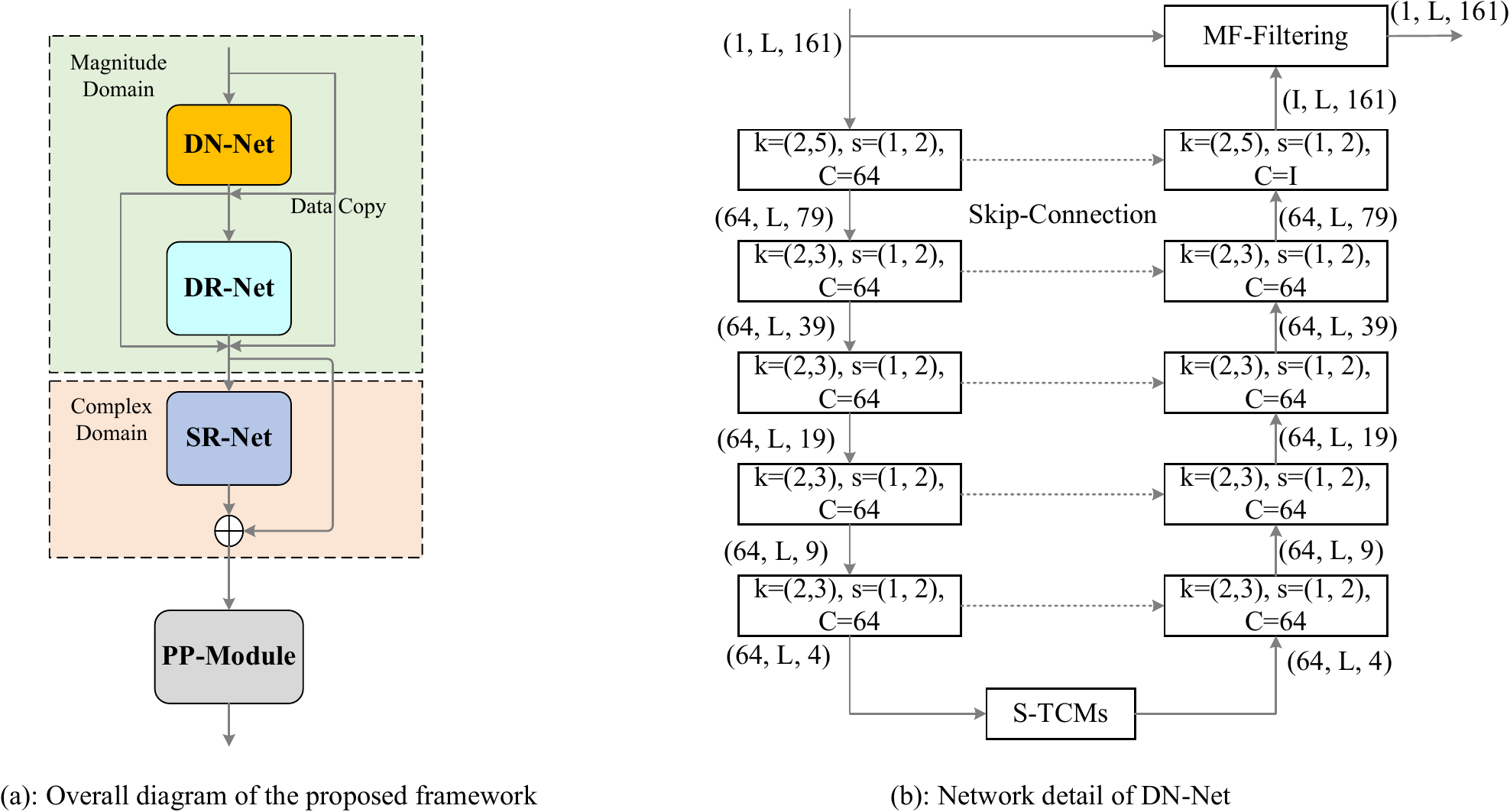}}
	\caption{Overall diagram of the proposed framework.}
	\label{fig:architecture}
	\vspace{-0.6cm}
\end{figure*}

From Eq.~({\ref{eqn3}}), the noise and late reverberation components can be gradually suppressed in a stagewise manner. Similar to~{\cite{li2020two}}, we decouple the complex spectrum optimization into a sequence learning problem \emph{w.r.t.} magntide and phase. Eq.~(\ref{eqn3}) can be changed into:
\begin{equation}
\label{eqn4}
p\left(X|Y\right) = p\left(|N| |Y\right)p\left(|R||Y, |N|\right)p\left(X|Y, |R|, |N| \right),
\end{equation}
where $| \cdot |$ denotes the modulo operator. One can find that we only obtain the estimated magnitudes in the first two stages and refine the complex spectrum of target in the third stage.
\vspace{-0.2cm}
\section{System description}
\label{system-overview}
\vspace{-0.2cm}
\subsection{Forward stream}
\label{forward-stream}
\vspace{-0.2cm}
The diagram of the proposed framework is shown in Figure~{\ref{fig:architecture}}(a). It consists of four parts, namely denoising module (DN-Net), dereverberation module (DR-Net), spectral refinement module (SR-Net), and post-processing (PP) module. Both DN-Net and DR-Net are operated in the magnitude domain, \emph{i.e.}, we first decouple the complex spectrum into magnitude and phase, and only the amplitude is processed, leaving the phase unchanged. After the second stage, we couple the processed magnitude and the original noisy phase to re-generate the coarse complex spectrum. For SR-Net, it receives both original and coarse complex spectra and further refines the spectrum details \emph{w.r.t.} magnitude and phase. Note that different from implicitly re-generating the spectrum from scratch, a global residual connection is introduced to enforce the network to only focus on the missing detail rather than the overall spectrum distribution. After SR-Net, despite the promising performance of previous stages, there still exist some residual noises in the non-active regions due to the nonlinear effect of DNNs and dataset mismatch, so a post-processing module is introduced to further suppress the distortion, which validates to benefit the subjective quality. In a nutshell, the whole procedure can be formulated as:
\begin{gather}
\label{eqn5}
|\tilde{X}^{dn}| = \mathcal{G}_{dn}\left( |Y|; \Phi_{1} \right),\\
|\tilde{X}^{dr}| = \mathcal{G}_{dr}\left( |\tilde{X}^{dn}|, |Y|; \Phi_{2} \right),\\
\tilde{X}^{dn} = |\tilde{X}^{dn}| \exp(j\theta_{Y}), \tilde{X}^{dr} = |\tilde{X}^{dr}| \exp(j\theta_{Y}),\\
\tilde{X}^{sr} = \tilde{X}^{dr} + \mathcal{G}_{sr}\left(\tilde{X}^{dn}, \tilde{X}^{dr}, Y; \Phi_{3} \right),\\
\tilde{X}^{pp} = \mathcal{G}_{pp} \left( \tilde{X}^{sr}; \Phi_{4} \right),
\end{gather}
where $\left( \tilde \cdot \right)^{dn}$, $\left( \tilde\cdot \right)^{dr}$, $ \left( \tilde\cdot \right)^{sr}$, $\left( \tilde \cdot \right)^{pp}$ denote the outputs of DN-Net, DR-Net, SR-Net and post-processing module, respectively. $\mathcal{G}_{dn}$, $\mathcal{G}_{dr}$, $\mathcal{G}_{sr}$, and $\mathcal{G}_{pp}$ are the functions of the corresponding four modules with parameter set $\Phi_{\left(\cdot\right)}$. $\theta_{Y}$ denotes the noisy phase.
\vspace{-0.4cm}
\subsection{Network configurations}
\label{network-configurations}
\vspace{-0.3cm}
Due to the promising performance of convolutional encoder-decoder topology in the SE applications~{\cite{braun2021towards, strake2020interspeech, zhao2020noisy}}, we employ it in the first three modules. As both DN-Net and DR-Net are operated in the magnitude domain, only one decoder is utilized to recover the magnitude. For SR-Net, similar to~{\cite{tan2019learning}}, we take two decoders to estimate both real and imaginary (RI) components. Take DN-Net as an example, the diagram of DN-Net is shown in Figure~{\ref{fig:architecture}}(b). The encoder is comprised of five blocks, each of which includes one convolutional layer, InstanceNorm (IN), and PReLU. For the decoder, it is the mirror version of the encoder, where each convolutional layer is replaced by the deconvolutional layer to recover the original size. Skip connection is utilized to mitigate the information loss during the mapping process. The kernel size, stride, and the number of channels are defined as $k$, $s$, and $C$, whose values are presented in the figure. The feature size in the $i${th} layer is formatted as $\left( C_{i}, L, F_{i}\right)$ in the channel, time, and frequency axis, respectively. One can find that we keep the time resolution unchanged, which guarantees the causal implementation. To capture the long-term temporal dependencies, we insert cascaded temporal convolutional modules (TCMs)~{\cite{luo2019conv}} in the bottleneck. To decrease the parameters, we opt to the squeezed version~{\cite{li2020two, li2021icassp}}, \emph{i.e.}, S-TCM, where the feature size is first compressed into 64 rather than 512 as the literature stated~{\cite{luo2019conv}}, followed by dilated convolutions. For each stage, we stack three groups of TCMs, each of which includes 6 S-TCMs with dilation rate $d = \{1, 2, 4, 8, 16, 32\}$.

For DN-Net and DR-Net, assuming the output of the last layer is denoted as $|\tilde{M}^{q}|\in \left( I, L, F \right), q\in\{dn, dr\}$. $I$ denotes the filter length and we set it as 5 in this paper. Then we conduct the multi-frame (MF) filtering to captilize on the correlations between neighboring frames~{\cite{mack2019deep}}, whose calculation gives as:
\vspace{-0.2cm}
\begin{gather}
\label{eqn6}
|\tilde{X}_{l, :}^{dn}| = \sum_{\tau=0}^{I-1} \tilde{M}^{dn}_{\tau, :, :} |Y_{l-\tau, :}|,\\ 
|\tilde{X}_{l, :}^{dr}| = \sum_{\tau=0}^{I-1} \tilde{M}^{dr}_{\tau, :, :} |\tilde{X}^{dn}_{l-\tau, :}|,
\end{gather}
\vspace{-0.8cm}
\subsection{Post-processing module}
\label{post-processing}
\vspace{-0.2cm}
With the powerful performance of the advanced network topology, the speech quality can be notably improved. Nonetheless, some residual noises still exist. The reason can be attributed as two-fold. Firstly, due to the data-driven property, the performance of DNNs is usually limited. Therefore, when the acoustic scenario is untrained, the performance will inevitably degrade, leading to some audible residual noises. Secondly, the network is usually highly nonlinear. In real environments, the unnatural distortion may arise due to the nonlinear effect.

Inspired by~{\cite{valin2018hybrid}}, we utilize an low-complexity approach as the post-processing module to further suppress the residual noise at the third stage output. Specifically, we adopt a light-weight network~{\cite{valin2018hybrid}} to obtain the gain function, which serves as the estimate of the speech presence probability (SPP) to recursively estimate the noise power spectral density (NPSD).  With the estimated NPSD, the MMSE-LSA estimator~{\cite{ephraim1985speech}} is leveraged to calculate the final gain and then applies to suppress the residual noise. In addition, we also adopt a cepstrum-based preprocessing scheme to suppress the harmonic components before estimating the NPSD~{\cite{hu2013cepstrum}}. As such, the over-estimation problem of NPSD can be mitigated in most cases.
\vspace{-0.2cm}
\section{Experiments}
\label{experiments}
\vspace{-0.2cm}
\subsection{Dataset}
\label{dataset}
\vspace{-0.2cm}
To evaluate the performance of our framework, we conduct extensive experiments on the DNS-Challenge dataset{\footnote{github.com/microsoft/DNS-Challenge/tree/master/datasets}}. It consists of a wide range of clean sets, noise clips, and RIRs, which simulate complicated acoustic scenarios. The clean set includes four categories, namely read speech, singing speech, emotional speech, and non-English speech. We find the quality \emph{w.r.t.} non-English is relatively poor, so we select parts of the utterances with high SNRs. Besides, we also include the Mandarin corpus from DiDiSpeech~{\cite{guo2020didispeech}} due to its high speech quality. The noise set includes over 60,000 clips and the total amount is around 181 hours. To formulate the real room scenarios, more than 115,000 RIRs are provided.

In this challenge, we totally generate around 2000 hours noisy-clean pairs. To be specific, we synthesize the noisy sets with around 1000 hours for read speech. For the Mandarin language, around 500 hours training sets are generated. The total duration \emph{w.r.t.} non-English, emotional, and singing utterances are about 400 hours, 50 hours, and 50 hours, respectively. During each mixing process, the clean speech is convolved with a randomly selected RIR, and then mix it with the noise vector under the SNR range of $\left(-5\rm{dB}, 15\rm{dB}\right)$. For training target, direct sound or early reflections (around 50ms) is usually chosen~{\cite{kinoshita2017neural, zhao2020monaural}}. However, we keep the first 100ms reflections in this study, which is based on two premises. Firstly, due to the performance limitation of DNNs, excessive dereverberation may cause speech discontinuity and hamper speech perception quality. Besides, despite dereverberation is not required in the challenge, we observe that moderate dereverberation can alleviate the smearing effect and improve the MOS. 

To compare the performance with other state-of-the-art (SOTA) SE systems, we also generate a relatively small training set (around 100 hours) from the DNS Challenge datasets. For testing, 50 noises are selected from MUSAN~{\cite{snyder2015musan}} and 100 RIRs are generated using the image method~{\cite{allen1979image}}. Five SNRs are evaluated, namely $\{-3\rm{dB}, 0\rm{dB}, 3\rm{dB}, 6\rm{dB}, 9\rm{dB}\}$ with 150 pairs in each case.
\vspace{-0.2cm}
\subsection{Training configurations}
\label{training-configurations}
\vspace{-0.2cm}
All the utterances are sampled at 16kHz and chunked to 8 seconds for training stability. The 20ms Hanning window is utilized, with 50\% overlap between adjacent frames. To extract the features, 320-point FFT is utilized to obtain the complex spectrum. Most recently, the effectiveness of power-compressed spectrum has been demonstrated in the dereverberation task~{\cite{li2021importance}}, so we conduct the compression toward the spectral magnitude before sending into the network, and the compression variable $\beta=0.5$, which is reported optimal. As the multi-stage paradigm is adopted, before training the network of current stage, we pre-train the network of the last stage and then freeze the weights. MSE is selected as the criterion to train the model with Adam optimizer~{\cite{kingma2014adam}}. The initialized learning rate (LR) is set to 0.001, and we halve the LR when validation loss does not increase for consecutive 2 epochs. The batch size is set to 32 at the utterance level.
\vspace{-0.2cm}
\subsection{Baselines}
\label{baselines}
\vspace{-0.2cm}
We compare the proposed framework with five advanced SE systems, namely GCRN~{\cite{tan2019learning}}, DCCRN~{\cite{hu2020dccrn}}, TSCN~{\cite{li2021icassp}}, AECNN~{\cite{pandey2019new}}, and Conv-TasNet~{\cite{luo2019conv}}. GCRN, DCCRN, and TSCN are complex-domain based approaches, which aim to recover both magnitude and phase information simultaneously. AECNN and Conv-TasNet belong to the time-domain family, where the network receives the raw waveform and then estimates the clean waveform directly. All the baselines are implemented with the best configurations mentioned in the literature. 
\vspace{-0.4cm}
\section{Results and Analysis}
\label{results-and-analysis}
\vspace{-0.2cm}
Three metrics are utilized to evaluate the performance of our framework, namely perceptual evaluation of speech quality (PESQ)~{\cite{rix2001perceptual}}, extended short-time objective intelligibility (ESTOI)~{\cite{jensen2016algorithm}}, and DNSMOS~{\cite{reddy2020dnsmos}}. PESQ, and ESTOI are to evaluate the objective performance of speech quality and intelligibility. DNSMOS is a non-intrusive perceptual objective metric, which accurately simulates the human subjective evaluation process.
\vspace{-0.2cm}
\subsection{Comparison with advanced baselines}
\label{comparison-with-advanced-baselines}
\vspace{-0.2cm}
The result comparison in terms of PESQ and ESTOI among different systems are shown in Tables~{\ref{tbl:pesq-comparison}} and~{\ref{tbl:estoi-comparison}}. ``Stage'' denotes the stage index of our approach and ``$\dagger$'' denotes the case when power compression is not adopted.  From the results, several observations can be made. Firstly, compared with time-domain-based models, better performance is achieved for complex-domain-based approaches. This indicates that it seems more suitable to perform simultaneous denoising and dereverberation in the T-F domain than the waveform level. Secondly, when the power compression is adopted before sending into the network, the performance can be notably improved. For example, in the first stage, when the range of the spectrum is compressed, it yields 0.11, and 1.38\% improvements in PESQ and ESTOI, respectively. This is because the power compression can decrease the dynamic range of the spectrum, which elevates the significance of low-energy regions. So the spectral details can be better recovered while some residual noises can also be suppressed. Thirdly, in the first three stages, both PESQ and ESTOI are gradually improved. However, when the PP is applied, both objective metrics are decreased. This is because despite the residual noise can be further suppressed, it also tends to cancel the speech components in the low-energy regions, which hampers the objective speech quality. Fourth, compared with previous systems, our model yields consistently better performance, showing the superiority of our system.
\vspace{-0.2cm}

\renewcommand\arraystretch{0.84}
\begin{table}[t]
	\caption{Objective results of different systems in terms of PESQ. $\textbf{BOLD}$ denotes the best result in each case. ``Cau.'' denotes whether the system is causal implementation.}
	\centering
	\large
	\resizebox{0.46\textwidth}{!}{
		\begin{tabular}{c|c|c|cccccc}
			\toprule
			Model &Stage &Cau. &-3\rm{dB} &0\rm{dB} &3\rm{dB} &6\rm{dB} &9\rm{dB} &Avg.\\
			\hline
			Noisy &- &- &1.62 &1.84 &1.95 &2.18 &2.36 &1.99\\
			\hline
			GCRN &- &$\checkmark$ &2.18 &2.41 &2.55 &2.76 &2.87 &2.55\\
			DCCRN &- &$\checkmark$ &2.21 &2.44 &2.58 &2.78 &2.92 &2.59 \\
			TSCN &- &$\checkmark$ &2.30 &2.50 &2.63 &2.82 &2.93 &2.63\\
			AECNN &- &$\times$ &2.09 &2.33 &2.45 &2.69 &2.79 &2.47\\
			Conv-TasNet &- &$\checkmark$ &2.06 &2.28 &2.39 &2.59 &2.72 &2.41\\
			\hline
			SDD-Net$^{\dagger}$(ours) &1 &$\checkmark$ &2.15 &2.31 &2.42 &2.57 &2.67 &2.42\\
			SDD-Net(ours) &1 &$\checkmark$ &2.26 &2.42 &2.52 &2.67 &2.77 &2.53\\
			SDD-Net(ours) &2 &$\checkmark$ &2.34 &2.55 &2.69 &2.87 &2.99 &2.69\\
			SDD-Net(ours) &3 &$\checkmark$ &\textbf{2.46} &\textbf{2.68} &\textbf{2.84} &\textbf{3.02} &\textbf{3.13} &\textbf{2.83}\\
			SDD-Net(ours) &4 &$\checkmark$ &2.32 &2.52 &2.68 &2.84 &2.98 &2.67\\
			\bottomrule
	\end{tabular}}
	\label{tbl:pesq-comparison}
	\vspace*{-0.39cm}
\end{table}

\renewcommand\arraystretch{0.80}
\begin{table}[t]
	\caption{Objective results of different systems in terms of ESTOI (in \%).}
	\centering
	\LARGE
	\resizebox{0.46\textwidth}{!}{
		\begin{tabular}{c|c|c|cccccc}
			\toprule
			Model &Stage &Cau. &-3\rm{dB} &0\rm{dB} &3\rm{dB} &6\rm{dB} &9\rm{dB} &Avg.\\
			\hline
			Noisy &- &- &46.28 &54.06 &60.65 &67.26 &73.19 &60.29\\
			\hline
			GCRN &- &$\checkmark$ &54.87 &63.52 &69.55 &75.53 &79.66 &68.63\\
			DCCRN &- &$\checkmark$ &55.93 &64.11 &70.16 &75.86 &80.46 &69.30 \\
			TSCN &- &$\checkmark$ &57.92 &65.55 &71.63 &77.15 &81.22 &70.69\\
			AECNN &- &$\times$ &51.92 &60.78 &66.95 &73.77 &77.87 &66.26\\
			Conv-TasNet &- &$\checkmark$ &51.16 &60.04 &66.29 &72.70 &77.40 &65.52\\
			\hline
			SDD-Net$^{\dagger}$(ours) &1 &$\checkmark$ &54.07 &61.78 &67.87 &73.46 &77.80 &66.99\\
			SDD-Net(ours) &1 &$\checkmark$ &56.58 &63.93 &69.73 &75.01 &78.86 &68.82\\
			SDD-Net(ours) &2 &$\checkmark$ &58.97 &66.71 &72.65 &77.80 &81.87 &71.60\\
			SDD-Net(ours) &3 &$\checkmark$ &\textbf{60.57} &\textbf{68.29} &\textbf{73.95} &\textbf{79.05} &\textbf{82.84} &\textbf{72.94}\\
			SDD-Net(ours) &4 &$\checkmark$ &60.50 &68.20 &73.79 &78.88 &82.60 &72.79\\
			\bottomrule
	\end{tabular}}
	\label{tbl:estoi-comparison}
	\vspace*{-0.66cm}
\end{table}

\begin{figure}[t]
	\centering
	\centerline{\includegraphics[width=0.85\columnwidth]{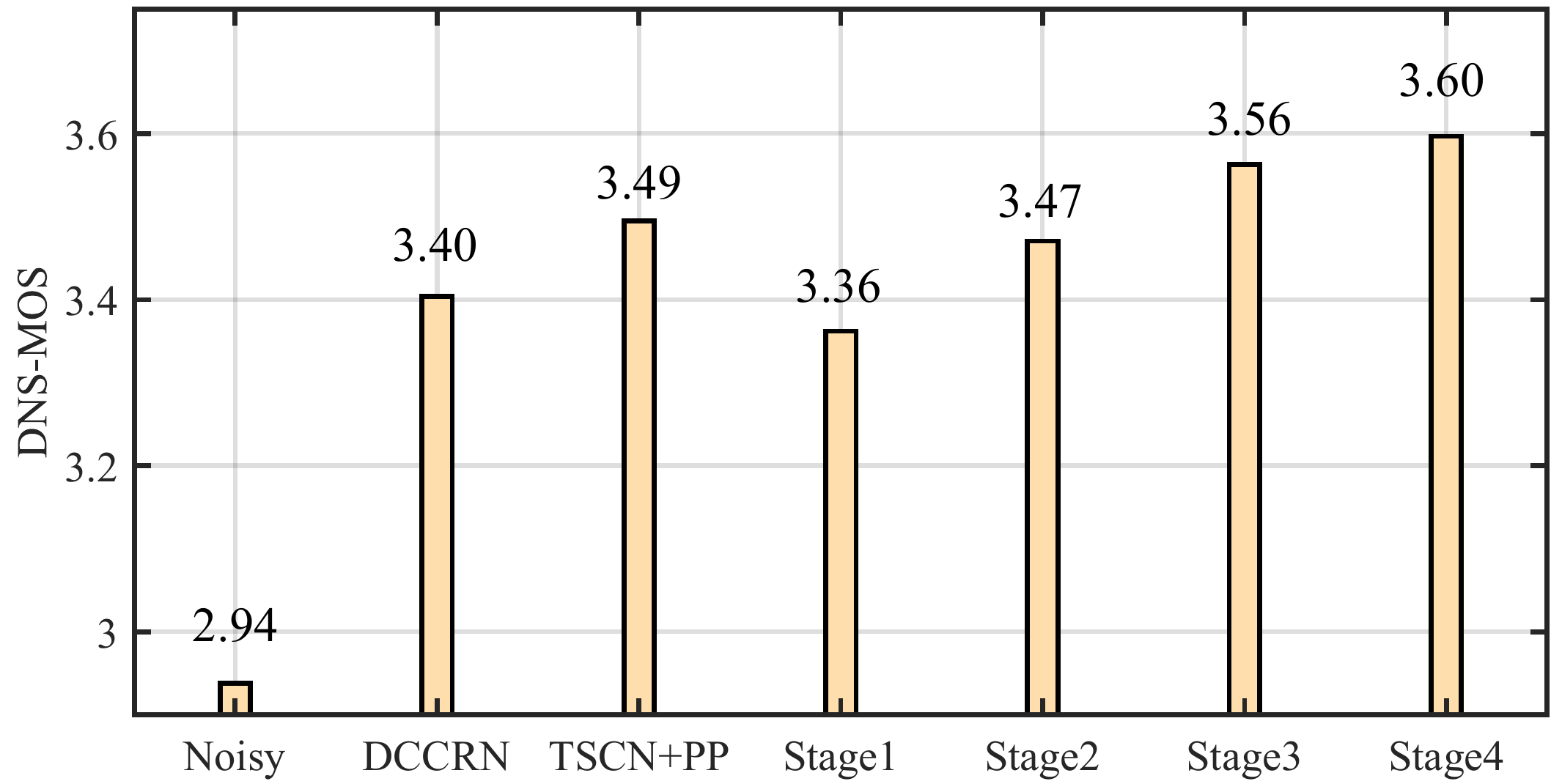}}
	\caption{DNSMOS comparison with previous champion schemes.}
	\label{fig:dnsmos}
	\vspace{-0.4cm}
\end{figure}
\vspace{-0.05cm}
\subsection{DNSMOS comparison with previous champions}
\label{dns-mos}
\vspace{-0.2cm}
We evaluate the average DNSMOS score of our approach on the released blind test set, which is shown in Figure~{\ref{fig:dnsmos}}. Besides, the scores of DCCRN and TSCN+PP~{\cite{li2021icassp}} are also provided, which won the first and second challenges, respectively. Several observations can be made. Firstly, compared with previous champion schemes, our system yields the highest score in the fourth stage, which shows the superiority of our method in improving the subjective perception quality. Secondly, despite the post-processing module decreases the objective metrics like PESQ and ESTOI, as stated in Section~{\ref{comparison-with-advanced-baselines}}, it can still improve the DNSMOS. It reveals the gap between objective metrics and human perception quality. Thirdly, in the previous challenges, the reverberation effect is usually neglected and most advanced schemes only concentrate on noise removal~{\cite{li2021icassp, hu2020dccrn}}. However, we find that when the dereverberation module is integrated into the framework, the overall DNSMOS score can be sizably improved. This phenomenon reveals the necessity and significance of the dereverberation task in improving human ratings. 

\renewcommand\arraystretch{1.10}
\begin{table}[t]
	\caption{Subjective evaluation with P.835 criterion on the DNS-Challenge.}
	\centering
	\Huge
	\resizebox{\columnwidth}{!}{
		\begin{tabular}{cc|cccccccc}
			\toprule
			\multicolumn{2}{c|}{Models} &Stationary &Emotional &Tonal &Non-English &Musical &English &Overall\\
			\hline
			
			\multirow{3}*{\rotatebox{90}{Speech}} &Noisy &4.02 &3.83 &\textbf{3.93} &3.80 &\textbf{3.97} &\textbf{3.87} &3.89\\
			&NSNet2(baseline) &3.53 &3.15 &3.44 &3.53 &3.23 &3.15 &3.35\\
			&SDD-Net &\textbf{4.03} &\textbf{3.90} &\textbf{3.93} &\textbf{3.96} &3.80 &3.76 &\textbf{3.90}\\
			\hline
			
			\multirow{3}*{\rotatebox{90}{Noise}} &Noisy &2.86 &1.93 &2.91 &3.11 &2.14 &2.30 &2.60\\
			&NSNet2(baseline) &4.21 &3.61 &4.23 &4.03 &3.11 &3.94 &3.88\\
			&SDD-Net &\textbf{4.78} &\textbf{4.66} &\textbf{4.73} &\textbf{4.65} &\textbf{4.53} &\textbf{4.61} &\textbf{4.65}\\
			\hline
			
			\multirow{3}*{\rotatebox{90}{Overall}} &Noisy &3.03 &2.28 &3.00 &3.04 &2.57 &2.52 &2.77\\
			&NSNet2(baseline) &3.28 &2.75 &3.31 &3.25 &2.78 &2.93 &3.07\\
			&SDD-Net &\textbf{3.92} &\textbf{3.79} &\textbf{3.79} &\textbf{3.84} &\textbf{3.71} &\textbf{3.63} &\textbf{3.78}\\
			\bottomrule
	\end{tabular}}
	\label{tbl:dns-challenge}
	\vspace*{-0.67cm}
\end{table}
\vspace{-0.2cm}
\subsection{Subjective results on the DNS-Challenge}
\label{subjective-results-on-the-dns}
\vspace{-0.2cm}
In Table~{\ref{tbl:dns-challenge}}, we present the subjective results of the submission with ITU-T P.835~{\cite{naderi2020crowdsourcing}}, which is provided by the organizer. Except for the overall subjective score, the standalone quality scores \emph{w.r.t.} speech distortion and noise removal also need to be considered. From the table, one can find that compared with noisy speech, where the speech components are distortionless, the baseline algorithm causes heavy speech distortion while our approach can even surpass the ideal case in some speech types. Besides, for denoising and overall conditions, our approach achieves dramatic improvements in MOS scores over the baseline. It indicates while our system can meet the requirement for background noise suppression in real acoustic scenarios, the speech components can also be well reserved.

In this study, the window size $T = 20$ms, with overlap $T_{s} = 10$ms. Therefore, the algorithm delay $T_{d} = 30$ms, which satisfies the latency requirement. In addition, the trainable parameter of the whole framework is 6.38 million, and the number of multiply-accumulate operations (MACS) is 60.07 million per frame. Therefore, the number of operations per second is 6.00 GMACS. We also evaluate the processing time, and the one frame processing time of the PyTorch implementation of the system is around 4.40ms on an Intel i5-4300U PC. 

\vspace{-0.1cm}
\section{Conclusions}
\label{conclusion}
\vspace{-0.1cm}
In this paper, we propose a novel framework to tackle the simultaneous denoising and dereverberation task. We adopt a chain optimization strategy and decompose the multi-target optimization in the complex domain into several stages. In the first two stages, we only focus on noise suppression and late reverberation removal in the magnitude domain.  Based on that, we can further refine the spectrum \emph{w.r.t.} magnitude and phase in the third stage by introducing a global residual connection. In addition, to further eliminate the residual noise, we adopt a post-processing module, which validates to be significant to improve the subjective quality. Experimental results show that our framework ranked first in MOS for the real-time track of the Interspeech 2021 DNS Challenge. 

\vspace{-0.1cm}
\section{Acknowledgement}
\label{acknowledgement}
\vspace{-0.1cm}
The authors would like to thank Y.~Hu at Northwestern Polytechnical University for providing the DNSMOS score of DCCRN in the blind test set.

\vfill\pagebreak
\bibliographystyle{IEEEtran}
\bibliography{refs}

\end{document}